\documentclass[twoside]{article}
\usepackage{fleqn,espcrc2}
\usepackage{epsfig}

\usepackage{graphicx}
\usepackage[figuresright]{rotating}

\newcommand{\AmS}{{\protect\the\textfont2
  A\kern-.1667em\lower.5ex\hbox{M}\kern-.125emS}}

\title{
\vspace*{-2cm}
\begin{flushright}
BI-TP2000/35
\end{flushright}
Deep inelastic scattering at low x: Generalized vector dominance and 
       the color dipole picture\thanks{Presented at Diffraction
       2000, Cetraro, Italy, September 2-7, 2000.}}

\author{D. Schildknecht\address{
        University of Bielefeld, Faculty for Physics,
        P.O. Box 10 01 31, D-33501 Bielefeld \\
        e-mail: Dieter.Schildknecht@physik.uni-bielefeld.de}
        \thanks{Supported by the BMBF, Berlin, Germany, Contract 05HT9PBA2} }

\begin{document}
\thispagestyle{empty}

\begin{abstract}
We summarize recent work on low-x deep inelastic scattering. The generalized
vector dominance/color-dipole picture (GVD/CDP) implies a scaling 
behavior for $\sigma_{\gamma^* p} (W^2, Q^2)\cong \sigma_{\gamma^* p} 
(\eta)$, with $\eta = (Q^2 + m^2_0)/\Lambda^2 (W^2)$ and yields an 
excellent representation of the experimental results on $\sigma_{\gamma^* p} 
(\eta)$. 

\vspace{1pc}
\end{abstract}

\maketitle

Two important observations \cite{1} were made on deep inelastic scattering 
(DIS) at low values of the Bjorken scaling variable
$x_{bj} \cong Q^2/W^2 \ll 1$, since HERA started running in 1993:

\medskip
i)
The diffractive production of high-mass states (of masses 
$M_X {\buildrel{<}\over{\sim}} 30$GeV) 
at an appreciable rate relative to the total virtual-photon proton cross 
section, $\sigma_{\gamma^* p} (W^2 , Q^2)$.
The sphericity and thrust analysis \cite{1} of the diffractively produced states 
revealed (approximate) agreement in shape with the final state found in 
$e^+ e^-$ annihilation at $\sqrt s = M_X$. This observation of high-mass 
diffractive production confirms the conceptual basis of generalized 
vector dominance (GVD) \cite{2}
that extends the role of the low-lying vector mesons in
photoproduction \cite{2a} to 
DIS at arbitrary $Q^2$, provided $x_{bj} \ll 1$. 

\medskip
ii)
An increase of $\sigma_{\gamma^* p} (W^2, Q^2)$ with increasing energy 
considerably stronger \cite{3}
than the smooth ``soft-pomeron'' behavior known from photoproduction 
and hadron-hadron scattering. This latter observation may have appeared to 
be unexpected from the point of view of GVD. A careful analysis, taking into 
account the quark-antiquark 
($q \bar q$) configuration in the $\gamma^* (q \bar q)$ transition, 
as well as two-gluon exchange as generic structure of the $(q \bar q)p$ 
interaction \cite{4,5,6}, (compare fig.1), however, 
reveals \cite{7}
that a stronger rise of $\sigma_{\gamma^* p} (W^2, Q^2)$ with energy 
than observed in photoproduction is entirely natural. In fact, one might 
have predicted a stronger rise than in photoproduction for $Q^2 \ge m^2_\rho$
in the generalized vector dominance/color dipole picture (GVD/CDP)
\cite{5,6,7}
prior to the experimental discovery. 

\noindent
\begin{figure}[ht]
\begin{center}
 \vskip -1cm
 \epsfig{file=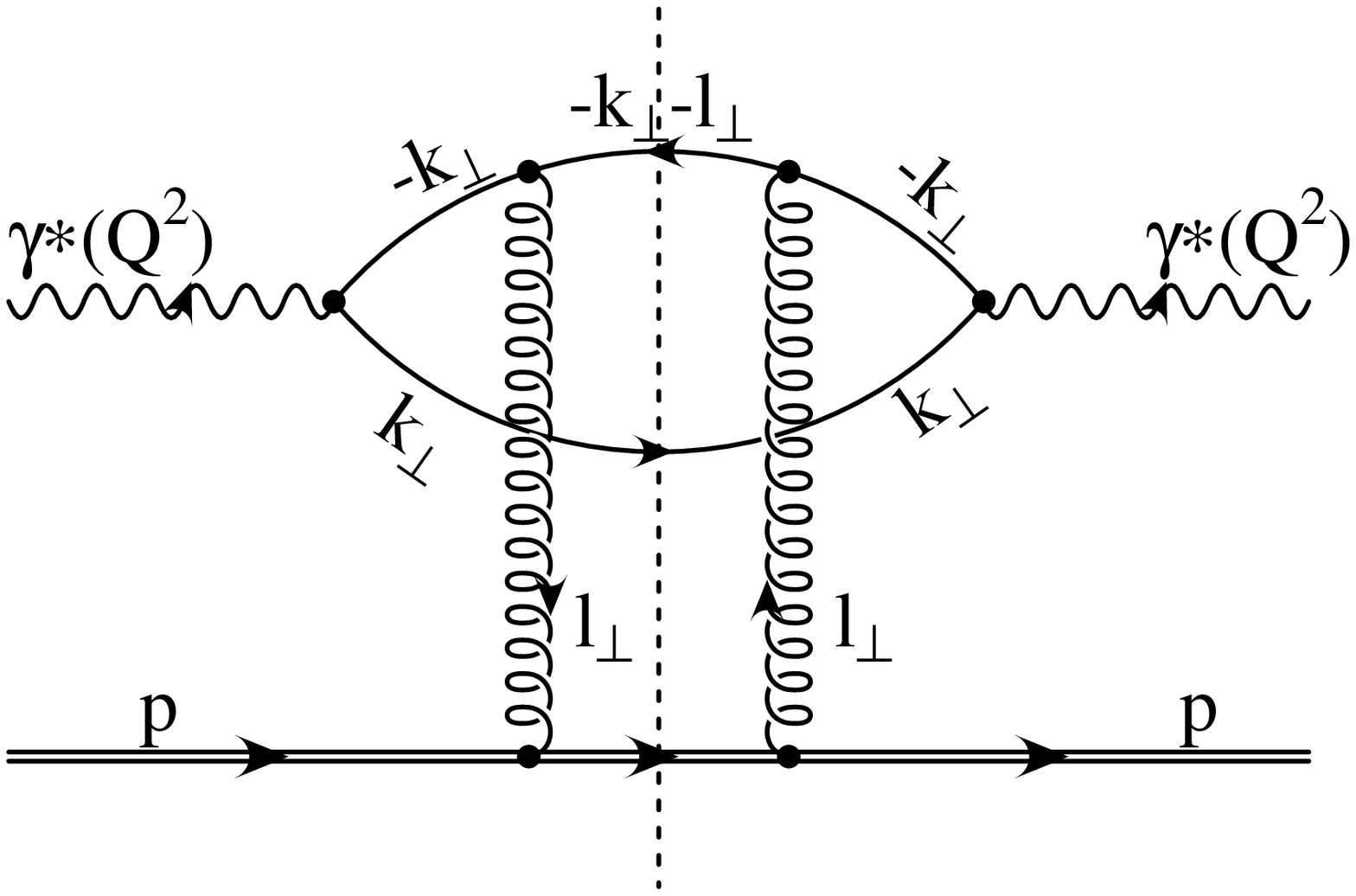,width=7.5cm}
 \vskip 1 mm
 \epsfig{file=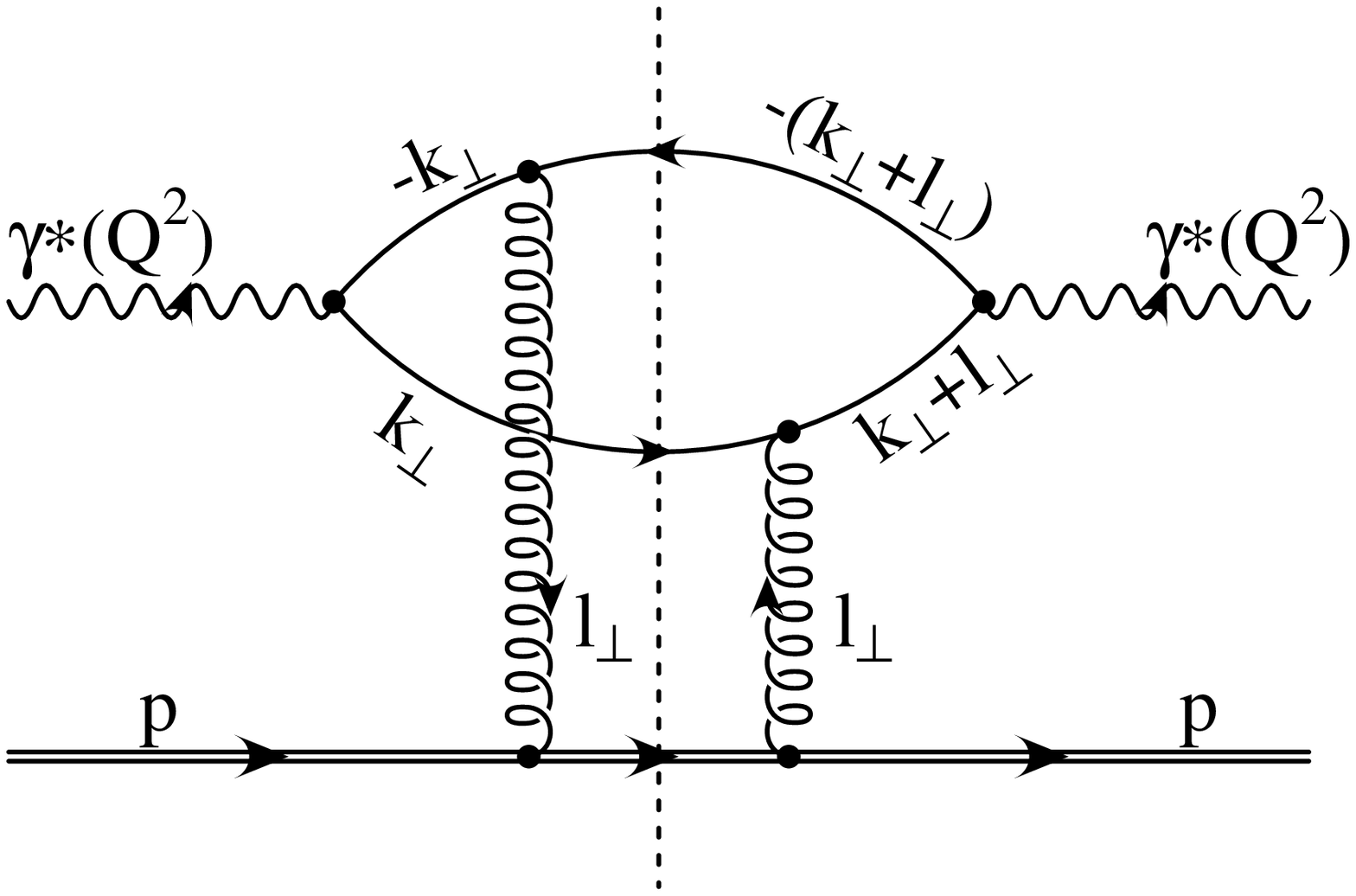,width=7.5cm}
 \vskip -1cm 
\end{center}
\caption{The two-gluon exchange.
\label{Fig1}}
\vskip -0.5cm
\end{figure}

We follow recent custom \cite{8,9,9a}
in employing the transverse-position-space representation as starting point, 
rather than proceeding in historic order \cite{5} from momentum space to 
transverse-position space \cite{6}. The representation 
\begin{eqnarray}
&&\sigma_{\gamma^* p}(W^2, Q^2) = \nonumber \\ 
&& \int d^2 r_\bot \int d z | \psi (r_\bot 
\sqrt{Q^2}, z, Q^2)|^2  \\
&&~~~~~~~~~~~~\times\sigma_{(q \bar q) p} (\vec r^{~2}_\bot , z, W^2) \nonumber
\end{eqnarray}
must be read \cite{6} in conjunction with 
\begin{eqnarray}
&&\sigma_{(q \bar q) p} (\vec r^{~2}_\bot , z , W^2)\nonumber \\
&&~~~~~=\int d^2 l_\bot 
\tilde\sigma (\vec l^{~2}_\bot, z, W^2) (1 - e^{-i \vec l_\bot \vec r_\bot}).
\end{eqnarray}
Inserting the ``color-dipole cross section'' $\sigma_{q \bar q)p} (\vec r^{~2}_\bot , 
z, W^2)$ from (2) into (1), together with the Fourier representation of the
``photon-wave-function'', $\psi (r_\bot \sqrt{Q^2}, z, Q^2)$, 
one indeed obtains \cite{6} 
the generic two-gluon exchange structure: the resulting representation of 
$\sigma_{\gamma^* p} (W^2, Q^2)$ is characterized by a linear combination of 
a diagonal and an off-diagonal term with respect to the masses of the 
ingoing and outgoing $q \bar q$ states that contribute with equal weight, 
but opposite in sign, to the virtual Compton forward-scattering 
amplitude. 

The form (2) of the color-dipole cross section (taking the limit of $r_\bot
\rightarrow \infty$) implies that the distribution in (the 
gluon-momentum-transfer variable) $\vec l_\bot$ should tend to zero 
sufficiently rapidly to yield a convergent result for the 
integral over $\tilde\sigma (\vec l^{~2}_\bot , z, W^2)$. One may think of 
introducing a Gaussian in $\vec l^{~2}_\bot$ for 
$\tilde\sigma (\vec l^{~2}_\bot ,z, W^2)$, 
but a $\delta$-function situated at a finite value of 
$\vec l^{~2}_\bot$ turns out to be equally successful as an effective description 
of the $\vec l^{~2}_\bot$ dependence of $\tilde\sigma (\vec l^{~2}_\bot , z, 
W^2)$, and, moreover, its consequences can be fully worked out analytically. 

The choice of \cite{7}
\begin{eqnarray}
&&\tilde\sigma_{(q \bar q)p} (\vec l^{~2}_\bot , z, W^2) \nonumber \\
&&~~~~=\sigma^{(\infty)} 
(W^2) \delta (\vec l^{~2}_\bot - z(1-z) \Lambda^2 (W^2)), 
\end{eqnarray}
when converted to $\vec r_\bot$ space according  to (2), explicitly realizes
\begin{itemize}
\item[i)]
color transparency, i.e.
\begin{eqnarray}
&&\!\!\!\!\!\!\!\!\!\! \sigma_{(q \bar q)p}  
(\vec r^{~2}_\bot , z, W^2) \rightarrow \Lambda^2 (W^2) z (1-z)
\vec r^{~2}_\bot, \nonumber \\
&&~~~~~~~{\rm for}\, z(1-z) r^{~2}_\bot \rightarrow 0,  
\end{eqnarray}
as well as 

\item[ii)]
unitarity, i.e.
\begin{eqnarray}
& &\sigma_{(q \bar q)p} (\vec r^{~2}_\bot , z, W^2) \rightarrow \sigma^{(\infty)}_{(q 
\bar q)p} (W^2), \nonumber \\
& &~~~~~~{\rm for}\, r_\bot \rightarrow \infty, 
\end{eqnarray}
where $\sigma^{(\infty)} (W^2)$ is to have a weak ``hadronlike''
energy dependence. 
\end{itemize}

\begin{figure}[ht]
\vspace*{-0.5cm}
\begin{center}
{\centerline{\epsfig{file=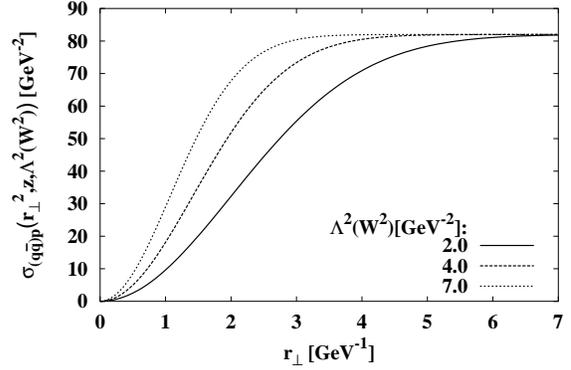,width=7.5cm}}}
\label{fig2}
\vspace*{-0.5cm}
\caption{Representation of the $(q\bar q)p$ - color dipole cross
section for realistic values of $\Lambda^2(W^2)$.}
\end{center}
\vspace*{-1cm}
\end{figure}

\noindent
Moreover, since the mass difference in off-diagonal transitions, $M_{q \bar q} 
\rightarrow M^\prime_{q \bar q} \not= M_{q \bar q}$, is determined by 
the magnitude of $\vec l^{~2}_\bot$, one expects that
\begin{itemize}
\item[iii)]
$\Lambda^2 (W^2)$ {\it increases} with the center-of-mass energy, $W$, 
rather than 
being a constant, independent of the energy of the $(q \bar q)p$ interaction.
\end{itemize}
The above properties i) to iii) 
of $\sigma_{(q \bar q)p} (\vec r^{~2}_\bot , z, W^2)$ are 
schematically depicted in fig.2, 
where for simplicity a Gaussian instead of a $\delta$-function was assumed, 
and the weakly $W$-dependent function $\sigma^{(\infty)}_{(q \bar q)p} (W^2)$,
was replaced by a constant. Figure 2
clearly displays the strong $W$ dependence, as $\Lambda^2(W^2)$, for $r_\bot 
\rightarrow 0$, and the weak one for $r_\bot \rightarrow \infty$. 

Taking into account
\begin{itemize}
\item[iv)]    
the dependence of  $|\psi|^2$ in (1) on 
 $\vec r^{~2}_\bot Q^2$,
\end{itemize}
we immediately conclude that with increasing $Q^2$, decreasing interquark 
separations become more dominant. As a consequence, the energy dependence 
of $\sigma_{\gamma^* p} (W^2, Q^2)$ becomes increasingly stronger with 
increasing $Q^2$, 
in agreement with the experimental data \cite{3}. 

In other words, the GVD/CDP which explicitly incorporates the
configuration of the $\gamma^*(q\bar q)$ transition, as well as the
generic two-gluon-exchange structure for the $(q\bar q)p$ interaction,
implies the striking change of the $W$ dependence with increasing
$Q^2$ observed experimentally.

Finally, dimensional analysis, in conjunction with an explicit evaluation of 
(1) in momentum space, reveals that, in good approximation, 

\begin{itemize}
\item[v)]
the dependence of 
$\sigma_{\gamma^* p}(W^2, Q^2)$ is determined by the
low-x scaling variable \cite{7}
\begin{equation}
\eta = \frac{Q^2 + m^2_0}{\Lambda^2(W^2)}, 
\end{equation}
where $m_0$ is a threshold mass $m_0 < m_\rho$.
\end{itemize}

Explicitly, 
\begin{eqnarray}
\sigma_{\gamma^*p} (W^2, Q^2)\!\!\! &=& \!\!\!\sigma_{\gamma p}(W^2) 
\frac{I (\eta , 
\frac{m^2_0}{\Lambda^2 (W^2)})}{I \left( \frac{m^2_0}{\Lambda^2 (W^2)},
\frac{m^2_0}{\Lambda^2(W^2)}\right)} \nonumber \\
&\cong&\!\!\! \sigma_{\gamma^* p} (\eta),
\end{eqnarray}
where $\sigma^{(\infty)}_{(q \bar q)p} (W^2)$ was substituted in terms of the 
photoproduction cross section, $\sigma_{\gamma p}(W^2)$. For 
details on (7) we refer to \cite{7}.
We only note the representation of the dominant transverse part of the 
function $I(\eta, \frac{m^2_0}{\Lambda^2(W^2)})$, 
\begin{eqnarray}
&&\!\!\!\!\! \!\!I^{(1)}_T \left( \eta , \frac{m^2_0}{\Lambda^2 (W^2)}\right) = 
\frac{1}{\pi}
\int\limits^\infty_{m^2_0} d M^2\!\!\!\! \int\limits^{(M + \Lambda (W^2))^2}_{(M - \Lambda(W^2))^2}
\!\!\!\!\!\!\! d M^{\prime 2} \nonumber  \\
&&\!\!\!\!\! \times \left\{ \frac{M^2 \pi \delta (M^2 - M^{\prime 2})}{(Q^2 + M^2)
(Q^2 + M^{\prime^2})}  \right. \\
&&\!\!\!\!\!\!\!\!\!\!\left. - \frac{(M^{\prime 2}\! - \! M^2\! -\! \Lambda^2
(W^2))\omega (M^2, M^{\prime 2}, \Lambda^2 (W^2))}
{2 (Q^2 + M^2)(Q^2+M^{\prime 2})} \right\}  , \nonumber
\end{eqnarray}
that displays the underlying structure of GVD and can be shown to 
approximately coincide with the ansatz of ``off-diagonal'' GVD \cite{10}
from the pre-QCD era. 
The explicit expression for $I (\eta , \frac{m^2_0}{\Lambda^2 (W^2)})$ 
in (7) is complicated, but simple results are obtained in the limits of 
small $\eta$, and of large $\eta$, respectively,
\begin{equation}
I \cong I(\eta) = \cases{ ln (1/\eta), & for $\eta\ll 1$, \cr 
                          \frac{1}{2\eta}, & for $\eta \gg 1$. \cr }
\end{equation}

\begin{figure}[ht]
\vspace*{-0.5cm}
\begin{center}
{\centerline{\epsfig{file=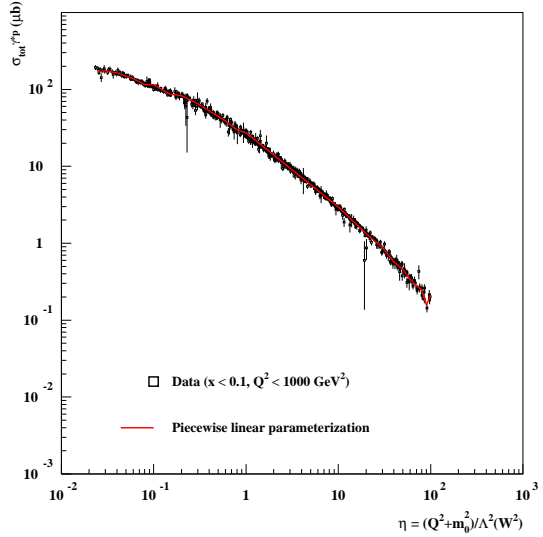,width=7.5cm}}}
\label{fig3}
\vspace*{-0.5cm}
\caption{The experimental data for $\sigma_{\gamma^* p} (W^2,Q^2)$ for 
$x \simeq Q^2/W^2 < 0.1$ vs. the low-x scaling variable 
$\eta = (Q^2 + m^2_0) / \Lambda^2 (W^2)$ (from ref.\protect\cite{7}).}
\end{center}
\vspace*{-1cm}
\end{figure}

An analysis \cite{7} of the experimental data \cite{3} reveals 

\medskip
i)
scaling in $\eta$ in a model-independent analysis (compare fig.3) 
that determines $m^2_0$ and
the functional behavior of $\Lambda^2 (W^2)$ in terms of three 
parameters, $\Lambda^2 (W^2) = C_1 (W^2 + W^2_0)^{C_2}$, and

\begin{figure}[!ht]
\vspace*{-1.5cm}
\begin{center}
{\centerline{\epsfig{file=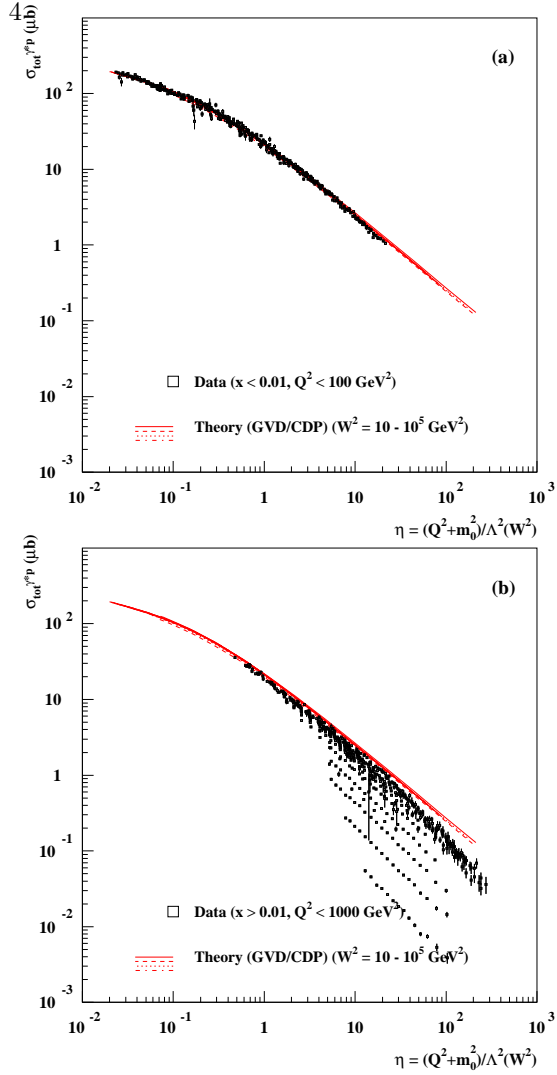,width=7.5cm}}}
\label{fig4}
\vspace*{-0.5cm}
\caption{The GVD/CDP scaling curve for $\sigma_{\gamma^* p}$ compared with the 
experimental data a) for $x < 0.01$, b) for $x > 0.01$ (from ref.\cite{7}).}
\end{center}
\vspace*{-1cm}
\end{figure}

\begin{figure}[!ht]
\vspace*{-1.5cm}
\begin{center}
{\centerline{\epsfig{file=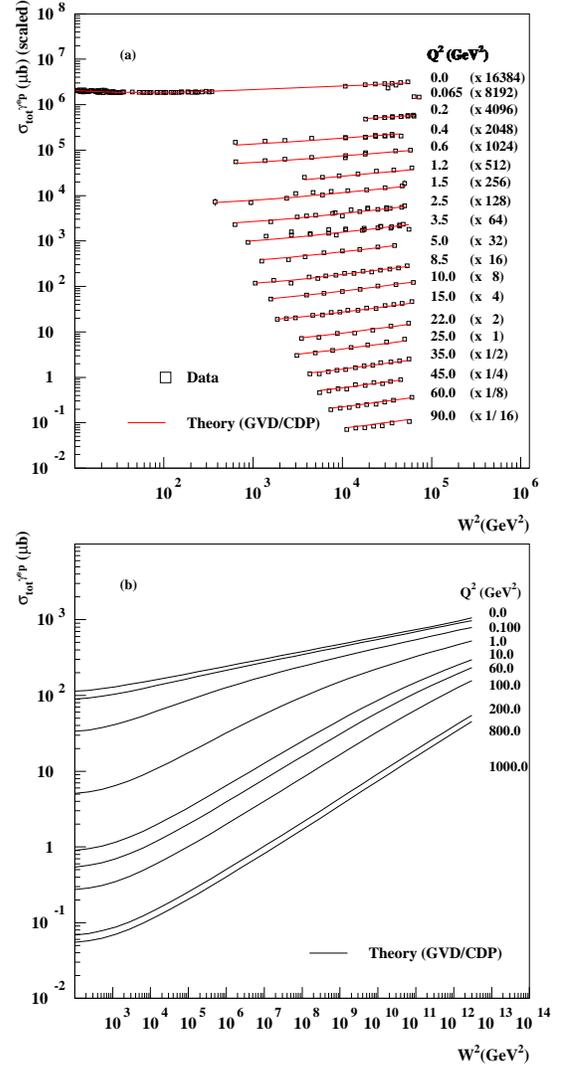,width=7.5cm}}}
\label{fig5}
\vspace*{-0.5cm}
\caption{The GVD/CDP predictions for $\sigma_{\gamma^* p}(W^2, Q^2)$ vs. $W^2$ 
at fixed $Q^2$\  a) in the presently accessable energy range
compared with experimental data for $x \le 0.01$, 
\  b) for asymptotic energies (from ref.\cite{7}).
}
\end{center}
\vspace*{-1cm}
\end{figure}

\medskip
ii)
good agreement of the $Q^2$ dependence, when $\Lambda^2(W^2)$ from the 
model-independent analysis is employed when evaluating $\sigma_{\gamma^* p}
(W^2, Q^2)$ in (7). 
We refer to \cite{7} 
for details and only show the results in figs. 4 and 5.

\medskip
In summary, a unique picture, the QCD-based generalized vector 
dominance/color-dipole picture (GVD/CDP) emerges for DIS at low $x_{bj}$ 
and any $Q^2$. The incoming virtual photon dissociates into a 
$q \bar q$-color-dipole state that propagates and undergoes diffractive 
forward scattering via an interaction of the generic structure of 
two-gluon exchange. 
The DIS experiments at low x measure the energy dependence of the 
$(q \bar q)$-color-dipole-proton interaction. With increasing 
$Q^2$, or, equivalently, decreasing transverse interquark separation, 
the generic 
two-gluon exchange structure of the $(q \bar q)$-color-dipole-proton 
interaction
implies the increasingly stronger $W$ dependence observed 
experimentally when $Q^2$ becomes large. 
At any $Q^2$, at
suffciently high energy, however, the $W$ dependence will settle down to the 
unitarity-preserving hadronic one.

\bigskip\noindent
{\bf ACKNOWLEDGEMENT}

\medskip
It is a pleasure to gratefully acknowledge the collaboration with 
Gorazd Cvetic, Arif Shoshi, Bernd Surrow and Mikhail Tentyukov that lead 
to the results reported here. I also thank the organizers of 
``Diffraction 2000'' in Cetraro, Italy, September 2 - 7, 2000,
for a very fruitful meeting in a 
magnificent setting and the participants, in particular John Dainton
and Alexei Kaidalow for useful discussions.

\end{document}